\documentclass[journal]{IEEEtran}

\usepackage{amsmath}
\usepackage{amssymb}
\usepackage{cite}
\usepackage{graphicx}
\usepackage{url}
\usepackage[hidelinks]{hyperref}

\begin{document}

\title{Structured Time–Frequency Representation for Channel Prediction}

\author{
Ning~Yang\textsuperscript{*},~\IEEEmembership{Senior Member,~IEEE},
Yuming~Xian,
Tong~Li,~\IEEEmembership{Senior Member,~IEEE},
Ning~Gao,~\IEEEmembership{Senior Member,~IEEE},
and Haijun~Zhang,~\IEEEmembership{Fellow,~IEEE}%
\thanks{Ning Yang is with the Institute of Automation,
Chinese Academy of Sciences, Beijing 100190, China
(e-mail: ning.yang@ia.ac.cn).}%
\thanks{Yuming Xian is with the School of Management Science
and Engineering, Central University of Finance and Economics,
Beijing 102206, China
(e-mail: 2024311756@email.cufe.edu.cn).}%
\thanks{Tong Li is with the College of Computer Science
and Electronic Engineering, Hunan University,
Changsha 410082, China
(e-mail: tliay@hnu.edu.cn).}%
\thanks{Ning Gao is with the School of Cyber Science
and Engineering, Southeast University, Nanjing, China
(e-mail: ninggao@seu.edu.cn).}%
\thanks{Haijun Zhang is with the School of Computer
and Communication Engineering, University of Science
and Technology Beijing, Beijing 100083, China
(e-mail: zhanghaijun@ustb.edu.cn).}%
\thanks{(*Corresponding author: Ning Yang.)}%
}

\maketitle

\begin{abstract}
Accurate CSI forecasting is essential for mitigating channel aging and supporting reliable downlink transmission in high-mobility wireless systems. In FDD systems, forecasting future DL CSI from historical UL observations requires joint temporal extrapolation and cross-band inference. However, defining historical UL as the input and future DL as the target does not specify how elapsed time and UL-to-DL band separation should be represented in a pretrained predictor's positional space. FAR-CSI establishes an FDD-aware representation of channel forecasting within a joint temporal and operating-band positional field. Within this field, historical UL observations and future DL targets occupy segregated regions according to their temporal and operating-band roles, preserving the distinction between observed UL context and unobserved DL prediction targets. This representation is instantiated through a unified structured routing rule that governs encoder positional assignment, decoder-field reconstruction, and future-DL target selection, ensuring the same FDD geometry is preserved from contextual encoding to prediction readout. Under matched backbone, initialization, and training conditions, FDD-aware routing reduces NMSE by 21.44\% relative to generic routing. Across the evaluated FDD baselines, FAR-CSI achieves higher CSI prediction accuracy and MRT-based spectral efficiency, while retaining its predictive advantage across mobility conditions, prediction horizons, reduced training-data regimes, and noisy historical observations.
\end{abstract}


% ============================================================
% I. INTRODUCTION
% ============================================================

\section{Introduction}
\label{sec:introduction}

\IEEEPARstart{T}{imely} channel state information (CSI) is essential for reliable downlink transmission and spatial precoding in mobile wireless systems. As user mobility increases, the channel can change between CSI acquisition and data transmission. This channel aging reduces the effectiveness of precoding based on outdated channel estimates \cite{R01_aging}. Acquiring CSI more frequently can alleviate aging, but increases estimation and signaling overhead. Channel forecasting offers another way to address this tradeoff by estimating the channel at the intended transmission instant from earlier observations \cite{R05_kalman,R06_fading}.

This challenge is more demanding in frequency-division duplex (FDD) systems. In time-division duplex (TDD) operation, calibrated channel reciprocity allows uplink (UL) measurements to provide information about the downlink (DL) channel at the same carrier frequency. In FDD operation, UL and DL use different frequency bands. Multipath delays produce different frequency-dependent phases in the two bands, so the complex UL channel measured at the base station cannot be directly substituted for the DL channel \cite{R07_extrapolation,R08_reciprocity}. Obtaining DL CSI conventionally requires downlink training and feedback, which introduce additional overhead and delay. Nevertheless, UL and DL signals can traverse much of the same scattering environment, so their multipath angles and delays may remain related across bands. Historical UL CSI therefore provides partial information about the unknown DL response. When DL CSI is needed at a future instant, this cross-band uncertainty is compounded by the channel's evolution during the intervening time \cite{R11_llm4cp,R13_stformer}. FDD channel prediction therefore couples temporal forecasting with cross-band inference.

Existing channel prediction methods have developed different approaches to this problem. Model-based approaches exploit propagation-domain structure to extrapolate channel states \cite{R02_pad,R34_st_ar_final}. Learning-based predictors capture temporal and spatial dependencies directly from CSI observations \cite{R03_transformer_final,R04_stnn}, while FDD-oriented methods learn cross-band relationships between UL inputs and DL targets \cite{R09_deepul2dl,R13_stformer}. These approaches provide ways to model temporal evolution and cross-band dependence jointly. They commonly specify the forecasting task through the available historical CSI and the required future CSI. This defines what is observed and what must be predicted, but leaves their internal positional organization to the chosen predictor.

More recently, pretrained models have introduced reusable
representations into channel prediction. LLM4CP adapts pretrained language-model representations to CSI forecasting \cite{R11_llm4cp}, while wireless foundation models such as WiFo and HeterCSI learn representations directly from CSI \cite{R18_wifo_final,R19_hetercsi}. These models strengthen the representation of channel content and its temporal, frequency, and spatial dependencies. In a pretrained patch-based CSI reconstruction predictor, positional indices are also part of the computation: they select positional embeddings during encoding and specify where observations and masked locations enter reconstruction. Adapting such a predictor to FDD forecasting therefore requires an explicit positional assignment for the historical UL observations and future DL targets. Specifying the input and output tensors alone does not establish that assignment.

The need for this assignment follows from the structure of the FDD task. Historical and future intervals intersect the UL and DL operating bands, forming a two-by-two time-band grid. Historical UL supplies the observations, and future DL defines the supervised target. Historical DL and future UL provide neither observations nor prediction targets, but they occupy the other two regions of this grid. A straightforward way to transfer the conventional input-to-output sequence formulation to a pretrained reconstruction predictor is to assign consecutive positions to the observed patches and target locations. Such compact indexing preserves sequence order but not necessarily the positions of those patches in the complete time-band grid. A future-DL target may then receive an address associated with a different time or band region, changing the positional embeddings and reconstruction locations used by the predictor. Reserving positions for historical DL and future UL keeps the observed and target patches at their designated grid addresses. In this way, a move forward in time and a move from UL to DL remain distinct in the positional layout. The reserved positions introduce no additional CSI measurements or supervised targets. The resulting challenge is to maintain this assignment consistently throughout prediction.

FAR-CSI addresses this challenge by formulating historical-UL-to-future-DL forecasting as structured conditional reconstruction over a joint time-band positional field. Its FDD-aware structured routing rule places the observed UL patches, future DL targets, and unavailable regions at their designated addresses. The same assignment governs encoder positional selection, masked decoder-field construction, and future-DL target readout. In this way, FAR-CSI establishes a consistent task representation throughout forecasting while using a pretrained CSI Transformer initialized from HeterCSI for channel modeling \cite{R19_hetercsi}.

The proposed representation is evaluated through matched
routing controls and comparisons with existing FDD predictors in terms of normalized mean-square error (NMSE) and
maximum-ratio-transmission (MRT)-based spectral efficiency.
Additional evaluations consider user mobility, forecasting horizon, limited training data, and noisy historical CSI.

The main contributions are summarized as follows:
\begin{itemize}

    \item An FDD-aware forecasting representation is introduced to organize historical UL observations and future DL targets within a joint temporal and operating-band field. It preserves designated positions for the observed context, prediction targets, and regions unavailable in the considered forecasting setting, without introducing additional measurements.

    \item A unified structured routing rule is developed to realize this representation in a pretrained CSI Transformer. It uses task-consistent positions for contextual encoding, masked reconstruction, and future-DL target selection.

    \item Matched routing experiments establish the predictive benefit of the complete FDD-aware configuration. Across the evaluated FDD baselines, FAR-CSI achieves the lowest NMSE and highest MRT-based spectral efficiency, with its accuracy advantage maintained across the tested conditions.

\end{itemize}

% ============================================================
% II. RELATED WORK
% ============================================================

\section{Related Work}
\label{sec:related_work}

\subsection{Model-Based Channel Prediction}

Model-based channel prediction derives future CSI from estimated channel dynamics or propagation parameters. Kalman filtering and spatio-temporal autoregressive models describe temporal evolution using state-space or autoregressive relationships \cite{R05_kalman,R34_st_ar_final}. Their accuracy depends on how well the estimated dynamics describe the channel over the prediction interval. Prony-based angular-delay prediction (PAD) takes a different approach by exploiting multipath structure in the angle, delay, and Doppler domains \cite{R02_pad}. This supports prediction under rapid channel variation, but requires sufficiently accurate recovery of propagation components. Limited antenna aperture, training bandwidth, or CSI quality can make those components difficult to resolve.

In FDD systems, channel prediction also involves a change of operating band. Frequency extrapolation uses shared propagation paths to infer DL responses from UL measurements \cite{R07_extrapolation}. Errors in the recovered path parameters become more consequential as the separation between UL and DL frequencies increases. Partial reciprocity-based methods combine UL-derived angle-delay-Doppler information with limited DL feedback for high-mobility FDD prediction \cite{R08_reciprocity}. They reduce dependence on full DL channel acquisition, but continue to require reliable parameter estimates, calibration, or supplementary DL information.

These approaches expose two different inference requirements in historical-UL-to-future-DL forecasting. Temporal prediction must account for channel evolution beyond the observation interval, whereas cross-band inference must account for the frequency-dependent difference between UL and DL responses. Learning from CSI can reduce reliance on explicit propagation-parameter recovery, but it does not remove the distinction between these two changes. Preserving that distinction within a single forecasting representation is a central requirement addressed by FAR-CSI.

\subsection{Task-Specific Neural Prediction}

Neural predictors learn channel relationships directly from CSI rather than requiring an explicit estimate of every propagation component. Early spatio-temporal networks combine antenna-domain features with temporal processing for channel prediction in massive multiple-input multiple-output (MIMO) systems using orthogonal frequency-division multiplexing (OFDM) \cite{R04_stnn}. Recurrent updates and local convolutional processing can make dependencies across distant snapshots difficult to capture. Transformer-based predictors address this modeling challenge through self-attention \cite{R37_attention}, allowing historical CSI snapshots to interact directly during multi-step prediction \cite{R03_transformer_final}. These advances improve temporal modeling, but same-band predictors do not by themselves establish the relationship between observations and targets at different operating frequencies.

FDD-oriented neural methods extend data-driven prediction across bands. Deep UL2DL learns a mapping from UL measurements to DL channel information \cite{R09_deepul2dl}, while deep transfer learning adapts learned channel knowledge to new FDD downlink conditions \cite{R30_transfer}. Cross-band mapping alone, however, does not describe how the target channel changes after the observed interval. CPcGAN considers time-varying TDD and FDD channel distributions through conditional adversarial learning \cite{R10_cpcgan}. STFormer more directly addresses the joint historical-UL-to-future-DL task. It combines a cross-time-frequency mapping with spatio-temporal attention, delay-Doppler feature fusion, and dimension-aware embeddings \cite{R13_stformer}. These methods establish the value of learning temporal and cross-band dependence together. Their principal design emphasis remains the learned mapping from available CSI to the desired output. Such a mapping need not explicitly assign positions to the historical DL and future UL regions that are unavailable during prediction.

Further work improves the neural operators used to learn these relationships. LinFormer replaces conventional attention with a time-aware multilayer perceptron to reduce the cost of MIMO channel prediction \cite{R35_linformer}. CPMamba and ChannelMamba use selective state-space modeling to process channel sequences under mobility \cite{R14_cpmamba,R15_shi}. Hypernetwork-based predictors adapt model parameters to evolving channel conditions \cite{R32_hyper}. These approaches address efficiency, long-range modeling, and adaptability. Improving a prediction operator, however, does not in itself determine how observed and unobserved time-band regions should be represented as distinct parts of the forecasting task.

Task-specific training also places demands on the available data. Without reusable CSI representations learned in advance, a predictor must learn both channel dependencies and the required UL-to-DL forecast from its downstream training samples. Fewer labeled FDD pairs leave less evidence for estimating these relationships \cite{R31_meta}, while noisy historical observations make the predictive information less reliable. These conditions can increase estimation difficulty and prediction error, motivating the use of pretrained channel knowledge. For FAR-CSI, the accompanying structural question is equally important: how to preserve the distinct time-band roles of the observed UL history and future DL target, including the regions without observations, when that knowledge is used for forecasting.

\subsection{Pretrained Representation Learning}

Pretraining learns reusable channel representations before adaptation to a particular prediction task. Masked autoencoding provides a natural approach by reconstructing missing content from visible observations \cite{R38_mae_final}. CSI-MAE and the Large Wireless Model (LWM) apply masked reconstruction to learn spatial and frequency-domain CSI relationships \cite{R23_csimae,R24_lwm}. LWM-Temporal further captures evolution across channel snapshots through sparse spatio-temporal attention \cite{R28_lwmtemporal}. BERT4MIMO, multi-task wireless foundation models, and WirelessGPT broaden pretrained modeling toward CSI reconstruction, generation, and wireless inference tasks \cite{R27_bert_final,R29_multitask,R25_wirelessgpt}. These approaches shift part of channel representation learning from task-specific labels to pretraining data. Forecasting nevertheless requires the learned representation to be connected to a particular set of available observations and prediction targets.

LLM4CP demonstrates this transfer for historical-UL-to-future-DL prediction using a pretrained language model \cite{R11_llm4cp}. It processes CSI in the frequency and delay domains and introduces channel-specific attention, embedding, and output modules around a GPT-2 backbone. This makes it possible to exploit pretrained sequence modeling for the FDD task. The adaptation, however, must bridge CSI structure and a backbone originally pretrained on language tokens. Its channel-specific modules encode wireless information, but the backbone's native sequence positions do not themselves specify the full historical/future and UL/DL arrangement, including the two regions without observations. The prediction is learned through the adapted input-to-output mapping rather than a common time-band reconstruction field with consistent observation and target addresses. This distinction motivates a CSI-native representation in which the FDD task structure is explicit throughout prediction.

Wireless-domain foundation models instead learn directly from CSI tensors. WiFo develops reusable representations for prediction across varying time-frequency configurations \cite{R18_wifo_final}. HeterCSI extends masked CSI pretraining to heterogeneous channel dimensions through three-dimensional time-frequency-antenna patches, scale-aware batching, and double masking \cite{R19_hetercsi}. These designs reduce the mismatch between the data modality and the pretrained model and provide structured channel features for downstream use. Yet supporting diverse CSI inputs is not the same as specifying where a fixed FDD forecasting task places its observations and supervised targets.

Recent work has strengthened wireless pretraining along two complementary directions. WiFo-2 addresses variation across system configurations and tasks by learning general representations from heterogeneous CSI \cite{R20_wifo2}. Broader coverage improves the reuse of channel knowledge, but it also makes accurate representation of the underlying CSI structure important. Adaptive 3D-RoPE addresses this issue through channel-adaptive positional encoding along the temporal, frequency, and antenna axes \cite{R21_rope3d}. AirFM-DDA takes a complementary route: it represents channels in the delay-Doppler-angle domain to expose multipath components and their dependencies during pretraining \cite{R22_airfm}. These advances enrich the coordinates or propagation features available to a predictor. The remaining FDD-specific question concerns how those features are assigned to the fixed observation and prediction regions of a cross-band forecasting task.

This issue is particularly clear in historical-UL-to-future-DL
forecasting. The available observations belong to the historical
interval and UL band, whereas the prediction targets belong to
the future interval and DL band. Pretrained CSI models provide
reusable channel representations, but these representations
do not by themselves specify how the two regions should be
organized for this forecasting task. FAR-CSI addresses this
remaining representation gap.

% ============================================================
% III. SYSTEM MODEL AND PROBLEM FORMULATION
% ============================================================

\section{System Model and Problem Formulation}
\label{sec:system_model}

\subsection{FDD Channel Representation}
\label{subsec:channel_representation}

Consider an FDD multicarrier link in which future DL CSI is predicted from historical UL CSI. Let $T_{\mathrm{h}}$ and $T_{\mathrm{f}}$ denote the numbers of historical and future CSI snapshots, respectively. The index
$t\in\{1,\ldots,T_{\mathrm{h}}+T_{\mathrm{f}}\}$
covers the complete historical and prediction interval. At time index $t$, the complex-valued channel associated with link direction $q$ is represented as
\begin{equation}
    \mathbf{H}_{t}^{q}
    \in
    \mathbb{C}^{P\times K\times N},
    \qquad
    q\in\{\mathrm{UL},\mathrm{DL}\},
    \label{eq:csi_snapshot}
\end{equation}
where $q$ identifies the UL or DL operating region. The dimensions $P$, $K$, and $N$ denote the number of polarization components, the number of frequency samples within the corresponding operating band, and the number of spatial elements per polarization, respectively. Thus, $\mathbf{H}_{t}^{\mathrm{UL}}$ and $\mathbf{H}_{t}^{\mathrm{DL}}$ share the same tensor organization while representing channel responses in distinct frequency bands.

The available channel observations consist exclusively of the first $T_{\mathrm{h}}$ UL snapshots,
\begin{equation}
    \mathcal{H}_{\mathrm{h}}^{\mathrm{UL}}
    =
    \left[
        \mathbf{H}_{1}^{\mathrm{UL}},
        \ldots,
        \mathbf{H}_{T_{\mathrm{h}}}^{\mathrm{UL}}
    \right]
    \in
    \mathbb{C}^{P\times T_{\mathrm{h}}\times K\times N},
    \label{eq:historical_ul}
\end{equation}
where $\mathcal{H}_{\mathrm{h}}^{\mathrm{UL}}$ denotes the historical UL CSI tensor. The desired prediction target contains the subsequent $T_{\mathrm{f}}$ DL snapshots,
\begin{equation}
    \mathcal{H}_{\mathrm{f}}^{\mathrm{DL}}
    =
    \left[
        \mathbf{H}_{T_{\mathrm{h}}+1}^{\mathrm{DL}},
        \ldots,
        \mathbf{H}_{T_{\mathrm{h}}+T_{\mathrm{f}}}^{\mathrm{DL}}
    \right]
    \in
    \mathbb{C}^{P\times T_{\mathrm{f}}\times K\times N},
    \label{eq:future_dl}
\end{equation}
where $\mathcal{H}_{\mathrm{f}}^{\mathrm{DL}}$ denotes the future DL CSI tensor. Equations~\eqref{eq:historical_ul} and~\eqref{eq:future_dl} therefore define two disjoint channel regions: the former is available at prediction time, whereas the latter must be inferred.

\subsection{FDD Channel Forecasting}
\label{subsec:fdd_forecasting}

Historical-UL-to-future-DL forecasting combines temporal extrapolation with cross-band inference. The input and output are separated not only in time but also in operating frequency, since the available CSI belongs to the historical UL region while the desired CSI belongs to the future DL region. Let
$\mathcal{F}_{\boldsymbol{\theta}}$
denote a channel-prediction function parameterized by
$\boldsymbol{\theta}$. The forecasting mapping is defined as
\begin{equation}
    \widehat{\mathcal{H}}_{\mathrm{f}}^{\mathrm{DL}}
    =
    \mathcal{F}_{\boldsymbol{\theta}}
    \left(
        \mathcal{H}_{\mathrm{h}}^{\mathrm{UL}}
    \right),
    \label{eq:fdd_mapping}
\end{equation}
where
$\widehat{\mathcal{H}}_{\mathrm{f}}^{\mathrm{DL}}
\in
\mathbb{C}^{P\times T_{\mathrm{f}}\times K\times N}$
denotes the predicted future DL CSI.

The mapping in Eq.~\eqref{eq:fdd_mapping} involves two coupled forms of channel dependence. Temporal dependence governs how the propagation channel evolves beyond the observed interval, whereas cross-band dependence relates the observed UL channel to the target DL channel across distinct operating frequencies. The forecasting problem is to determine the predictor parameters such that both dependencies are recovered accurately from historical UL observations.

\subsection{Forecasting Objective}
\label{subsec:prediction_metric}

Let
\begin{equation}
    \mathcal{D}
    =
    \left\{
        \left(
            \mathcal{H}_{\mathrm{h},b}^{\mathrm{UL}},
            \mathcal{H}_{\mathrm{f},b}^{\mathrm{DL}}
        \right)
    \right\}_{b=1}^{B}
    \label{eq:forecasting_dataset}
\end{equation}
denote a set of $B$ paired forecasting samples, where
$b\in\{1,\ldots,B\}$
indexes an individual channel sample. The desired predictor minimizes the normalized reconstruction error over the future DL channel tensors. Accordingly, the forecasting objective is formulated as
\begin{equation}
    \boldsymbol{\theta}^{\star}
    =
    \arg\min_{\boldsymbol{\theta}}
    \frac{
        \displaystyle
        \sum_{b=1}^{B}
        \left\|
            \mathcal{F}_{\boldsymbol{\theta}}
            \left(
                \mathcal{H}_{\mathrm{h},b}^{\mathrm{UL}}
            \right)
            -
            \mathcal{H}_{\mathrm{f},b}^{\mathrm{DL}}
        \right\|_{\mathrm{F}}^{2}
    }{
        \displaystyle
        \sum_{b=1}^{B}
        \left\|
            \mathcal{H}_{\mathrm{f},b}^{\mathrm{DL}}
        \right\|_{\mathrm{F}}^{2}
    },
    \label{eq:fdd_objective}
\end{equation}
where $\|\cdot\|_{\mathrm{F}}$ denotes the Frobenius norm and $\boldsymbol{\theta}^{\star}$ denotes the optimized predictor parameters. The numerator accumulates the complex-valued reconstruction error over all future-DL coefficients, while the denominator normalizes this error by the corresponding target-channel energy. This ratio-of-sums formulation measures the aggregate prediction error relative to the total target-channel energy, providing an energy-weighted evaluation of future-DL CSI reconstruction.

For any evaluation set $\mathcal{S}$, the same criterion is reported as NMSE,
\begin{equation}
    \mathrm{NMSE}(\mathcal{S})
    =
    \frac{
        \displaystyle
        \sum_{b\in\mathcal{S}}
        \left\|
            \widehat{\mathcal{H}}_{\mathrm{f},b}^{\mathrm{DL}}
            -
            \mathcal{H}_{\mathrm{f},b}^{\mathrm{DL}}
        \right\|_{\mathrm{F}}^{2}
    }{
        \displaystyle
        \sum_{b\in\mathcal{S}}
        \left\|
            \mathcal{H}_{\mathrm{f},b}^{\mathrm{DL}}
        \right\|_{\mathrm{F}}^{2}
    }.
    \label{eq:ros_nmse}
\end{equation}

A lower NMSE indicates that the predicted tensor more closely reproduces the complex-valued future DL channel.

% ============================================================
% IV. FAR-CSI
% ============================================================
\begin{figure*}[!t]
\centering
\includegraphics[width=\textwidth]{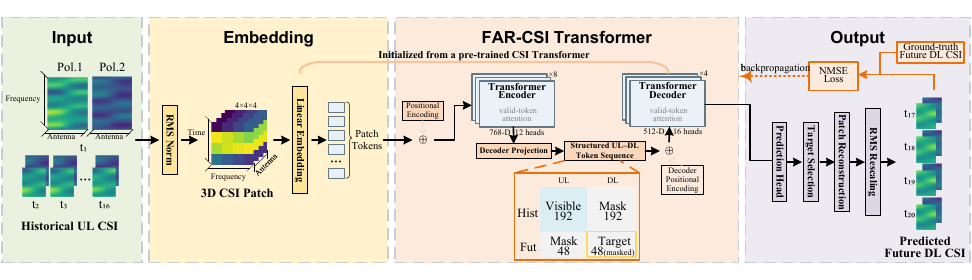}
\caption{Overall architecture of FAR-CSI for historical-UL-to-future-DL forecasting. Historical UL CSI is normalized and partitioned into local time-frequency-antenna patches. FDD-aware structured routing then assigns the observed UL tokens and unobserved channel regions to a joint temporal and operating-band positional field. Contextual representations of the observations are inserted into their designated decoder positions, while all remaining positions are mask-initialized. Reconstruction outputs located in the future-DL target region are selected, unpatched, and restored to the physical CSI amplitude scale.}
\label{fig:farcsi_framework}
\end{figure*}

\section{FAR-CSI}
\label{sec:far_csi}

\subsection{Framework Overview}
\label{subsec:far_overview}

FAR-CSI formulates historical-UL-to-future-DL forecasting as
structured reconstruction over a joint time-band positional field.
The field distinguishes historical from future time and UL from DL
operating bands. Historical UL provides the observed context, future
DL defines the prediction target, and the other two time-band regions
retain their positions without supplying channel measurements.

As illustrated in Fig.~\ref{fig:farcsi_framework}, the framework has
three stages. First, historical UL CSI is normalized and divided into
local time-frequency-antenna patches to obtain channel-content tokens.
Second, FDD-aware structured routing assigns these observations and
all unobserved locations to a common positional field. The assigned
addresses govern both encoder positional embeddings and the placement
of contextualized observations in the decoder. Third, the decoder
reconstructs the field, and the outputs at future-DL target positions
are selected and converted back to complex-valued CSI. The routing
therefore maintains the same time-band organization from encoding to
prediction readout.

A CSI Transformer initialized from HeterCSI \cite{R19_hetercsi}
provides the pretrained channel-modeling backbone. FAR-CSI defines
how the FDD forecasting task is represented and routed through this
backbone.

\subsection{CSI Patch Representation}
\label{subsec:csi_tokenization}

The first stage converts the available historical UL CSI into patch
representations. Normalization uses the observed UL history alone,
and its scale factors are retained to recover the predicted channel
in the original amplitude domain.

For sample $b$, the stabilized root-mean-square (RMS) scale of the
observed history is
\begin{equation}
    \rho_b
    =
    \max\!\left\{
        \sqrt{
            \frac{
                \left\|\mathcal{H}_{\mathrm{h},b}^{\mathrm{UL}}\right\|_{\mathrm{F}}^2
            }{P T_{\mathrm{h}} K N}
        },\,
        \epsilon
    \right\},
    \label{eq:sample_rms}
\end{equation}
where $\epsilon=10^{-12}$ is a numerical safeguard. Let
$\mathcal{H}_{\mathrm{h},b,p}^{\mathrm{UL}}\in
\mathbb{C}^{T_{\mathrm{h}}\times K\times N}$ denote the
$p$th polarization slice of the observed history, with
$p\in\{1,\ldots,P\}$. Its normalization scale is
\begin{equation}
    s_{b,p}
    =
    \rho_b
    \sqrt{
        \max\!\left\{
            \frac{
                \left\|
                    \mathcal{H}_{\mathrm{h},b,p}^{\mathrm{UL}}/\rho_b
                \right\|_{\mathrm{F}}^2
            }{T_{\mathrm{h}} K N},\,
            \epsilon
        \right\}
    }.
    \label{eq:polarization_rms}
\end{equation}
The normalized historical UL stream is
\begin{equation}
    \widetilde{\mathcal{H}}_{\mathrm{h},b,p}^{\mathrm{UL}}
    =
    \frac{\mathcal{H}_{\mathrm{h},b,p}^{\mathrm{UL}}}{s_{b,p}}.
    \label{eq:polarization_normalization}
\end{equation}
This operation standardizes the input scale separately for each
polarization while retaining $s_{b,p}$ for amplitude restoration.
Polarization streams are processed with shared predictor parameters
and regrouped after reconstruction.

Let $q_t$, $q_f$, and $q_a$ denote the patch sizes along the time,
frequency, and antenna dimensions, respectively. The historical UL
stream for each polarization, of size
$T_{\mathrm{h}}\times K\times N$, is partitioned into non-overlapping
$q_t\times q_f\times q_a$ patches. Assuming that $q_t$ divides both $T_{\mathrm{h}}$ and $T_{\mathrm{f}}$,
and that $q_f$ and $q_a$ divide $K$ and $N$, respectively, the
number of observed patches is
\begin{equation}
    N_{\mathrm{obs}}
    =
    \frac{T_{\mathrm{h}}}{q_t}
    \frac{K}{q_f}
    \frac{N}{q_a}.
    \label{eq:num_observed_patches}
\end{equation}

Let $\mathbf{P}_i\in\mathbb{C}^{q_t\times q_f\times q_a}$ denote the $i$th observed patch, with $i\in\{1,\ldots,N_{\mathrm{obs}}\}$, and define $Q=q_tq_fq_a$ as its number of complex coefficients. The real and imaginary parts are vectorized and concatenated:
\begin{equation}
    \mathbf{x}_i
    =
    \begin{bmatrix}
        \operatorname{vec}\!\left(\Re\{\mathbf{P}_i\}\right)\\
        \operatorname{vec}\!\left(\Im\{\mathbf{P}_i\}\right)
    \end{bmatrix}
    \in\mathbb{R}^{2Q}.
    \label{eq:patch_vector}
\end{equation}
The patch embedding operator produces the encoder content token
\begin{equation}
    \mathbf{e}_i
    =
    \mathcal{E}_{\mathrm{patch}}(\mathbf{x}_i)
    \in\mathbb{R}^{d_{\mathrm{enc}}},
    \label{eq:patch_embedding}
\end{equation}
where $d_{\mathrm{enc}}$ is the encoder embedding dimension.
Patch embeddings describe observed CSI content; the routing rule
introduced next assigns each token its position in the complete
FDD forecasting field.

\subsection{FDD-Aware Structured Routing}
\label{subsec:fdd_routing}

\subsubsection{Joint Time-Band Forecasting Field}

The routing field provides a single address space for the observed
UL history, the future DL targets, and the unobserved historical-DL
and future-UL regions. This organization is important because the
forecast crosses two distinct boundaries: the end of the observed
time interval and the separation between UL and DL operating bands.
The two unobserved regions occupy positions in the field, but they
contribute no CSI measurements and receive no direct target
supervision. Keeping their addresses allows the observation and
target tokens to be located according to the same time-band layout.

Using the patch sizes defined in Section~\ref{subsec:csi_tokenization},
let the numbers of historical and future temporal blocks be
$N_t^{\mathrm{h}}=T_{\mathrm{h}}/q_t$ and
$N_t^{\mathrm{f}}=T_{\mathrm{f}}/q_t$.
Let $N_f^{\mathrm{UL}}$ and $N_f^{\mathrm{DL}}$ denote the
frequency-block counts in the two operating bands, with
$N_f^{\mathrm{UL}}=N_f^{\mathrm{DL}}=K/q_f$ for the channel
representation in Section~\ref{subsec:channel_representation}.
The number of antenna blocks is $N_a=N/q_a$.
Define the total temporal and frequency-block counts by
$N_t=N_t^{\mathrm{h}}+N_t^{\mathrm{f}}$ and
$N_f=N_f^{\mathrm{UL}}+N_f^{\mathrm{DL}}$.
The full positional field contains
\begin{equation}
    N_{\mathrm{field}}=N_tN_fN_a
    =
    \left(N_t^{\mathrm{h}}+N_t^{\mathrm{f}}\right)
    \left(N_f^{\mathrm{UL}}+N_f^{\mathrm{DL}}\right)N_a.
    \label{eq:field_size}
\end{equation}
This field is the address space used for masked reconstruction.
Only historical-UL positions contain observations; the decoder
initializes every other position with a mask token. Future-DL
positions are the designated prediction targets.

\subsubsection{Observation and Target Assignment}

Each patch location is identified by a temporal-block index $i_t$,
a frequency-block index $i_f$, and an antenna-block index $i_a$.
Using zero-based indices,
\begin{equation}
    \begin{aligned}
        0&\le i_t<N_t,\\
        0&\le i_f<N_f,\\
        0&\le i_a<N_a.
    \end{aligned}
    \label{eq:routing_indices}
\end{equation}
The historical interval occupies
$0\le i_t<N_t^{\mathrm{h}}$ and the future interval occupies
$N_t^{\mathrm{h}}\le i_t<N_t$.
Similarly, the UL band occupies
$0\le i_f<N_f^{\mathrm{UL}}$ and the DL band occupies
$N_f^{\mathrm{UL}}\le i_f<N_f$.
These coordinates are converted to the one-dimensional addresses
used by the Transformer through
\begin{equation}
    \pi(i_t,i_f,i_a)
    =
    (i_tN_f+i_f)N_a+i_a,
    \label{eq:global_position}
\end{equation}
where $0\le\pi(i_t,i_f,i_a)<N_{\mathrm{field}}$.
Thus, the address of a patch depends on its position in the full
time-band field, rather than its order among the observed tokens.
A change of temporal block alters $i_t$, while a change of
operating band alters $i_f$; the two transitions thus retain
distinct offsets in the discrete address field. These addresses
index the positional-embedding tables of the adopted pretrained
CSI Transformer, whose available positions cover the required
forecasting field.

The historical-UL observation positions form the set
\begin{equation}
    \begin{aligned}
    \Omega_{\mathrm{obs}}=\big\{\pi(i_t,i_f,i_a)\,\big|\;&
        0\le i_t<N_t^{\mathrm{h}},\\
        &0\le i_f<N_f^{\mathrm{UL}},\\
        &0\le i_a<N_a\big\},
    \end{aligned}
    \label{eq:omega_obs}
\end{equation}
with cardinality
$|\Omega_{\mathrm{obs}}|=N_t^{\mathrm{h}}N_f^{\mathrm{UL}}N_a
=N_{\mathrm{obs}}$.
The future-DL target positions are
\begin{equation}
    \begin{aligned}
    \Omega_{\mathrm{tgt}}=\big\{\pi(i_t,i_f,i_a)\,\big|\;&
        N_t^{\mathrm{h}}\le i_t<N_t,\\
        &N_f^{\mathrm{UL}}\le i_f<N_f,\\
        &0\le i_a<N_a\big\},
    \end{aligned}
    \label{eq:omega_tgt}
\end{equation}
where
$|\Omega_{\mathrm{tgt}}|=N_t^{\mathrm{f}}N_f^{\mathrm{DL}}N_a$.
The two sets are disjoint and use the same global indexing rule.
In contrast, assigning observed and target patches to consecutive
sequence positions does not generally preserve their designated
addresses in this time-band field, even if the decoder has the
same total number of positions. Consequently, the routing changes
which pretrained positional embeddings the encoder uses and where
the decoder places observed tokens and prediction targets.

Define the remaining, non-target masked positions as
\begin{equation}
    \Omega_{\mathrm{aux}}
    =
    \{0,\ldots,N_{\mathrm{field}}-1\}
    \setminus(\Omega_{\mathrm{obs}}\cup\Omega_{\mathrm{tgt}}).
    \label{eq:omega_aux}
\end{equation}
These positions correspond to historical DL and future UL.
Their number is
\begin{equation}
    |\Omega_{\mathrm{aux}}|
    =
    \left(
        N_t^{\mathrm{h}}N_f^{\mathrm{DL}}
        +N_t^{\mathrm{f}}N_f^{\mathrm{UL}}
    \right)N_a.
    \label{eq:num_aux_positions}
\end{equation}
All positions outside $\Omega_{\mathrm{obs}}$ are mask-initialized,
but the loss is applied to $\Omega_{\mathrm{tgt}}$ alone.
The auxiliary positions preserve the complete positional layout
without introducing additional channel observations or prediction
targets.

\subsubsection{Consistent Routing Across Prediction}

FAR-CSI uses the same observation and target addresses throughout
the prediction process. During encoding, the content token
$\mathbf{e}_i$ receives the encoder positional embedding at its
assigned address $o_i\in\Omega_{\mathrm{obs}}$.
During decoder construction, the corresponding contextualized
token is placed at address $o_i$, while all other addresses are
initialized with mask tokens. The decoder positional embeddings
are then added over the complete field. Finally, reconstructed
patches at $\Omega_{\mathrm{tgt}}$ are selected as the future DL
prediction.

The assignment is therefore part of the computation, not a label
attached after prediction. It determines the encoder positions,
the arrangement of the decoder inputs, and the target readout
under one shared time-band indexing rule. This is the complete
FDD-aware routing operation evaluated in the matched comparison
of Section~\ref{subsec:routing_ablation}.

\subsection{Structured Contextual Reconstruction}
\label{subsec:contextual_prediction}

Let
\begin{equation}
    \mathbf{o}=(o_1,\ldots,o_{N_{\mathrm{obs}}})
    \label{eq:ordered_observation_positions}
\end{equation}
be an ordered enumeration of $\Omega_{\mathrm{obs}}$.
Each patch-content token $\mathbf{e}_i$ is combined with the
encoder positional embedding inherited from the pretrained backbone,
$\mathbf{p}^{\mathrm{enc}}_{o_i}\in\mathbb{R}^{d_{\mathrm{enc}}}$.
The encoder jointly contextualizes the observed tokens, and a
learned projection maps the outputs into the decoder space.
Denote the projected contextual token at position $o_i$ by
$\mathbf{z}_i\in\mathbb{R}^{d_{\mathrm{dec}}}$, where
$d_{\mathrm{dec}}$ is the decoder embedding dimension.

Let $\mathbf{m}\in\mathbb{R}^{d_{\mathrm{dec}}}$ be the learned
mask token, and let
$\mathbf{p}^{\mathrm{dec}}_r\in\mathbb{R}^{d_{\mathrm{dec}}}$
denote the decoder positional embedding at address
$0\le r<N_{\mathrm{field}}$.
The decoder input at each location is
\begin{equation}
    \mathbf{u}^{(0)}_r
    =
    \begin{cases}
        \mathbf{z}_i+\mathbf{p}^{\mathrm{dec}}_r,
        & r=o_i,\\[3pt]
        \mathbf{m}+\mathbf{p}^{\mathrm{dec}}_r,
        & r\notin\Omega_{\mathrm{obs}}.
    \end{cases}
    \label{eq:decoder_field}
\end{equation}
Here $i$ is uniquely determined by $r=o_i$ because the
observation addresses are distinct. The complete field includes
both the supervised future-DL targets and the remaining masked
positions defined in Section~\ref{subsec:fdd_routing}.

After contextual reconstruction, the prediction head converts
each decoder output at $r\in\Omega_{\mathrm{tgt}}$ into a real
vector of length $2Q$. Its real and imaginary components are
combined to recover a complex patch of size
$q_t\times q_f\times q_a$.
The target patches are assembled in their temporal, frequency,
and antenna order to form the normalized future DL CSI for each
polarization. Restoring the observed-history scale $s_{b,p}$
and regrouping polarization streams yields
\begin{equation}
    \widehat{\mathcal{H}}_{\mathrm{f},b}^{\mathrm{DL}}
    \in\mathbb{C}^{P\times T_{\mathrm{f}}\times K\times N}.
    \label{eq:reconstructed_future_dl}
\end{equation}
This output has the same dimensions and organization as the
target in Eq.~\eqref{eq:future_dl}, completing the forecasting
mapping in Eq.~\eqref{eq:fdd_mapping}.

\subsection{Training Objective}
\label{subsec:initialization_training}

The CSI Transformer is initialized from a pretrained HeterCSI
checkpoint \cite{R19_hetercsi}, and its trainable parameters are
fine-tuned for FDD forecasting. Supervision is applied to the
future-DL target set $\Omega_{\mathrm{tgt}}$.
The forecasting objective in Section~\ref{subsec:prediction_metric}
is optimized through the physical-domain ratio-of-sums (RoS)
loss over each effective training batch $\mathcal{B}$:
\begin{equation}
    \mathcal{L}_{\mathrm{RoS}}
    =
    \frac{
        \displaystyle
        \sum_{b\in\mathcal{B}}
        \left\|
            \widehat{\mathcal{H}}_{\mathrm{f},b}^{\mathrm{DL}}
            -\mathcal{H}_{\mathrm{f},b}^{\mathrm{DL}}
        \right\|_{\mathrm{F}}^2
    }{
        \displaystyle
        \sum_{b\in\mathcal{B}}
        \left\|
            \mathcal{H}_{\mathrm{f},b}^{\mathrm{DL}}
        \right\|_{\mathrm{F}}^2
    }.
    \label{eq:training_loss}
\end{equation}
The reconstructed CSI is restored to its original amplitude
scale before computing this loss. The other masked locations
contribute to the decoder field but receive no direct
reconstruction supervision. Details of the backbone and
optimization settings are given in Section~\ref{subsec:experimental_setup}.

% ============================================================
% V. EXPERIMENTS
% ============================================================

\section{Experiments}
\label{sec:experiments}

\subsection{Experimental Setup}
\label{subsec:experimental_setup}

\subsubsection{Dataset}
The experiments use the channel-prediction benchmark released with LLM4CP \cite{R11_llm4cp}, which was generated using QuaDRiGa \cite{R43_quadriga} under the Third Generation Partnership Project (3GPP) urban macro (UMa) non-line-of-sight (NLOS) setting \cite{R44_tr38901}. The base station employs a dual-polarized $4\times4$ uniform planar array (UPA), while the user equipment (UE) has a single antenna. The UL and DL channels each span $8.64$~MHz over $K=48$ resource blocks (RBs), with a pilot-frequency interval of $180$~kHz. The UL center frequency is $2.4$~GHz, and the evaluated FDD benchmark uses adjacent UL and DL frequency regions. Consecutive channel snapshots are separated by $0.5$~ms. Each forecasting instance uses $16$ historical UL snapshots to predict four future snapshots. The primary FDD task predicts future DL CSI, whereas the complementary TDD evaluation considers same-band forecasting from historical UL CSI to future UL CSI.

For the present evaluation, the available $9{,}000$ development samples are divided using a predefined $9{:}1$ split, yielding $8{,}100$ training samples and $900$ validation samples with disjoint sample identities. The separate test set contains $10{,}000$ samples covering ten velocities from $10$ to $100$~km/h, with $1{,}000$ samples at each velocity. The validation and test sample identities remain unchanged across the subsequent evaluations, while the reduced-data experiments draw smaller subsets from the original training split.

\subsubsection{Model and Training Settings}

FAR-CSI is initialized from the pretrained HeterCSI CSI
Transformer \cite{R19_hetercsi}, and all model parameters
are fine-tuned for the downstream FDD forecasting task.
The backbone uses $q_t=q_f=q_a=4$ CSI patches, an
eight-block encoder with $d_{\mathrm{enc}}=768$ and $12$
attention heads, and a four-block decoder with
$d_{\mathrm{dec}}=512$ and $16$ attention heads.

For the evaluated configuration, $T_{\mathrm{h}}=16$,
$T_{\mathrm{f}}=4$, $K=48$, $N=16$, and $P=2$.
The joint field contains $N_t^{\mathrm{h}}=4$ historical
and $N_t^{\mathrm{f}}=1$ future temporal blocks,
$N_f^{\mathrm{UL}}=N_f^{\mathrm{DL}}=12$ frequency blocks,
and $N_a=4$ antenna blocks per polarization.
Accordingly, $N_{\mathrm{field}}=480$ positions comprise
$N_{\mathrm{obs}}=192$ observed-UL positions,
$|\Omega_{\mathrm{tgt}}|=48$ future-DL target positions,
and $|\Omega_{\mathrm{aux}}|=240$ other masked positions.
The patch output dimension is $2Q=128$.

FAR-CSI is trained for $30$ epochs in 32-bit floating point
using AdamW \cite{R45_adamw} with an initial learning rate
of $10^{-4}$, weight decay of $0.05$, an effective batch
size of $16$ formed from microbatches of $4$, two warm-up
epochs followed by cosine learning-rate decay to a minimum
learning-rate ratio of $0.01$, and gradient clipping at $1$.
For the RoS training loss, the denominator is evaluated
over the entire effective batch, while the numerator
contributions are accumulated across microbatches before
the optimizer update.

Three independent training runs are conducted, and the
checkpoint with the lowest validation ratio-of-sums NMSE
is selected from each run for final evaluation.

\begin{table*}[!t]
    \caption{Clean-input overall prediction and communication performance.
    For methods evaluated over three independent runs, the reported value
    is the mean across runs. The best value within each duplex setting is
    shown in bold.}
    \label{tab:overall_performance}
    \centering
    \footnotesize
    \setlength{\tabcolsep}{1pt}
    \renewcommand{\arraystretch}{1.12}

    \textbf{FDD}\\[1mm]
    \begin{tabular*}{\textwidth}{
        @{\extracolsep{\fill}}
        lccccccccc
        @{}
    }
        \hline
        Metric
        & UL Persistence
        & RNN
        & GRU
        & LSTM
        & CNN
        & Transformer
        & LLM4CP
        & WiFo
        & FAR-CSI \\
        \hline
        NMSE
        & 1.9283
        & 0.5287
        & 0.6019
        & 0.6486
        & 0.9602
        & 0.4639
        & 0.3816
        & 1.1285
        & $\mathbf{0.1502}$ \\
        SE (bit/s/Hz)
        & 6.0754
        & 6.1167
        & 5.9942
        & 5.9857
        & 5.0802
        & 6.3553
        & 6.6251
        & 3.7870
        & $\mathbf{7.1054}$ \\
        \hline
    \end{tabular*}

    \vspace{2mm}

    \textbf{TDD}\\[1mm]
    \begin{tabular*}{\textwidth}{
        @{\extracolsep{\fill}}
        lccccccccc
        @{}
    }
        \hline
        Metric
        & No Prediction
        & RNN
        & GRU
        & LSTM
        & CNN
        & Transformer
        & LLM4CP
        & PAD3
        & Generic CSI Transformer \\
        \hline
        NMSE
        & 0.6931
        & 0.0907
        & 0.0457
        & 0.0554
        & 0.0227
        & 0.0304
        & 0.0136
        & $\mathbf{6.57\times10^{-7}}$
        & 0.001475 \\
        SE (bit/s/Hz)
        & 7.2440
        & 7.1218
        & 7.1822
        & 7.1748
        & 7.2453
        & 7.2668
        & 7.2625
        & $\mathbf{7.3309}$
        & 7.32654 \\
        \hline
    \end{tabular*}
\end{table*}

\subsubsection{Baselines}
The FDD comparison includes UL Persistence, recurrent neural network (RNN), gated recurrent unit (GRU), long short-term memory (LSTM), convolutional neural network (CNN), Transformer, LLM4CP, WiFo, and FAR-CSI. For TDD, the comparison includes No Prediction, RNN, GRU, LSTM, CNN, Transformer, LLM4CP, PAD3, and the Generic CSI Transformer. The Generic CSI Transformer employs the same pretrained CSI backbone and downstream training policy with generic routing for same-band forecasting. PAD3 \cite{R02_pad} serves as a deterministic model-based TDD reference.

The reimplemented Transformer is trained from scratch using Adam \cite{R46_adam} with a learning rate of $10^{-4}$, weight decay of $10^{-4}$, and an effective batch size of $32$ for at most $500$ epochs; it uses a constant learning rate without warm-up, early stopping with a patience of $30$ epochs, and best-validation checkpoint selection. Its training objective is the physical-domain ratio-of-sums NMSE. RNN, GRU, LSTM, CNN, and LLM4CP use their released task checkpoints with the corresponding preprocessing \cite{R11_llm4cp}, while WiFo \cite{R18_wifo_final} uses pretrained weights with the evaluated FDD masking adapter and no downstream optimizer update.

FAR-CSI, the reimplemented Transformer, and the Generic CSI Transformer are each evaluated over three independent training runs. Table~\ref{tab:overall_performance} reports the mean across these runs, whereas fixed released checkpoints and deterministic references are reported as single values. The matched routing control in Section~\ref{subsec:routing_ablation} reports mean $\pm$ standard deviation across three paired runs.

\subsubsection{Evaluation Protocol}
Prediction accuracy is evaluated using the physical-domain ratio-of-sums NMSE defined in Section~\ref{subsec:prediction_metric}. Communication utility is additionally assessed using MRT at a communication signal-to-noise ratio (SNR) of $10$~dB. Let $\mathcal{R}$ denote the set of evaluated resources, indexed by test sample, polarization component, future time step, and frequency resource. For each $j\in\mathcal{R}$, let $\mathbf{h}_j,\widehat{\mathbf{h}}_j\in\mathbb{C}^{16}$ denote the true and predicted spatial channels, respectively. The unit-norm MRT beamformer is
\begin{equation}
    \mathbf{w}_j
    =
    \begin{cases}
        \widehat{\mathbf{h}}_j/
        \left\|\widehat{\mathbf{h}}_j\right\|_2,
        &
        \left\|\widehat{\mathbf{h}}_j\right\|_2>0,\\[2pt]
        \mathbf{0},
        &
        \text{otherwise},
    \end{cases}
    \label{eq:mrt_beamformer}
\end{equation}
and the noise variance is referenced to the true mean spatial-channel power as
\begin{equation}
    \sigma_j^2
    =
    \frac{\left\|\mathbf{h}_j\right\|_2^2}{16}
    10^{-10/10}.
    \label{eq:se_noise_variance}
\end{equation}
The MRT-based spectral efficiency (SE) is
\begin{equation}
    \mathrm{SE}
    =
    \frac{1}{|\mathcal{R}|}
    \sum_{j\in\mathcal{R}}
    \log_2
    \left(
        1+
        \frac{
            \left|
                \mathbf{w}_j^{H}\mathbf{h}_j
            \right|^2
        }{
            \sigma_j^2
        }
    \right),
    \label{eq:mrt_se}
\end{equation}

The data-efficiency and noise-robustness experiments follow two additional protocols.
The reduced-data experiments use nested $5\%$, $10\%$, $25\%$, $50\%$, and $100\%$ subsets of the original training split while retaining the same validation and test sets. For noisy-input evaluation, the trained checkpoints are kept fixed and circular complex Gaussian noise is added only to the historical CSI input at SNR levels of $0$, $5$, $10$, $15$, $20$, and $25$~dB, while the future CSI target remains clean. The noise power is scaled according to the measured power of each historical input to target the specified SNR. For each test sample and SNR condition, the noise realization is fixed across methods so that all predictors receive the same corrupted input. The clean-input condition is evaluated separately from the finite-SNR settings.

% ============================================================
% V-B. Overall Prediction Performance
% ============================================================

\subsection{Overall Prediction Performance}
\label{subsec:overall_performance}

Table~\ref{tab:overall_performance} compares the evaluated methods under the clean-input setting. In FDD forecasting, FAR-CSI achieves the lowest NMSE and the highest MRT-based SE. Relative to LLM4CP and the reimplemented Transformer, it reduces NMSE by $60.6\%$ and $67.6\%$, respectively. The higher MRT-based SE further shows that the reduction in CSI prediction error translates into improved beamforming performance.

Under TDD forecasting, PAD3 achieves the strongest clean-input performance, while the Generic CSI Transformer is the strongest evaluated learning-based predictor. The contrast with the FDD results highlights the additional difficulty introduced when prediction spans both time and operating frequency.

% ============================================================
% V-C. Effect of FDD-Aware Structured Routing
% ============================================================

\subsection{Effect of FDD-Aware Structured Routing}
\label{subsec:routing_ablation}

To evaluate the contribution of FDD-aware structured routing, FAR-CSI is compared with a matched Generic Routing configuration using the same pretrained CSI Transformer, initialization, trainable modules, data splits, and downstream training policy. Generic Routing assigns the 192 historical-UL observation patches to consecutive positions 0--191 and the 48 future-DL targets to positions 192--239 within a 480-position decoder field. FAR-CSI instead uses the observation and target positions defined by $\Omega_{\mathrm{obs}}$ and $\Omega_{\mathrm{tgt}}$ in Section~\ref{subsec:fdd_routing}. Both configurations retain the same decoder size and are trained for 30 epochs using three paired random seeds. For each configuration and seed, the checkpoint with the lowest validation ratio-of-sums NMSE is selected for evaluation on the clean FDD test set.

\begin{table}[!t]
    \caption{Matched FDD routing ablation on the clean test set. Each run uses its best-validation checkpoint. NMSE values are reported as mean $\pm$ standard deviation over three paired seeds.}
    \label{tab:routing_ablation}
    \centering
    \begin{tabular}{lc}
        \hline
        Routing Strategy & NMSE \\
        \hline
        Generic Routing & $0.19134\pm0.00631$ \\
        FDD-Aware Routing & $\mathbf{0.15025\pm0.00223}$ \\
        \hline
    \end{tabular}
\end{table}

As shown in Table~\ref{tab:routing_ablation}, FDD-aware routing yields a mean paired relative NMSE reduction of $21.44\%\pm2.08\%$, with the same direction of improvement in all three paired runs. This controlled comparison supports the value of organizing CSI observations and prediction targets according to their joint temporal and operating-band roles. With the pretrained backbone and training protocol held fixed, this comparison establishes the additional predictive benefit of the complete FDD-aware routing configuration. The results support task-consistent positional organization as an effective means of adapting pretrained CSI reconstruction to historical-UL-to-future-DL forecasting.

% ============================================================
% V-D. Performance Across Mobility and Forecast Horizons
% ============================================================

\subsection{Performance Across Mobility and Forecast Horizons}
\label{subsec:mobility_horizon}

\begin{figure}[!t]
    \centering
    \includegraphics[width=\columnwidth]{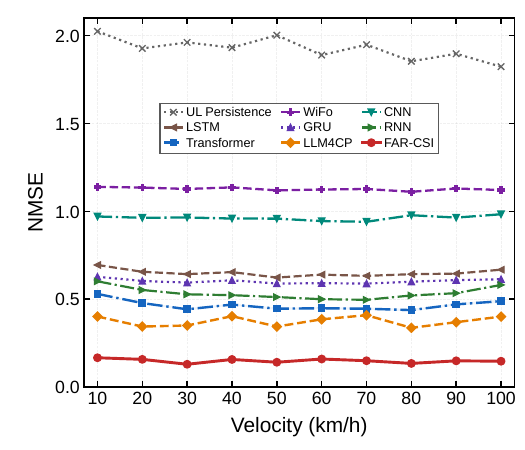}
    \caption{FDD NMSE across user velocities from $10$ to $100$~km/h under clean-input evaluation.}
    \label{fig:fdd_velocity}
\end{figure}

Across the evaluated mobility range, FAR-CSI maintains the lowest NMSE among the FDD methods, as shown in Fig.~\ref{fig:fdd_velocity}. The advantage therefore extends across the evaluated mobility range rather than being confined to a particular speed.

\begin{figure}[!t]
    \centering
    \includegraphics[width=\columnwidth]{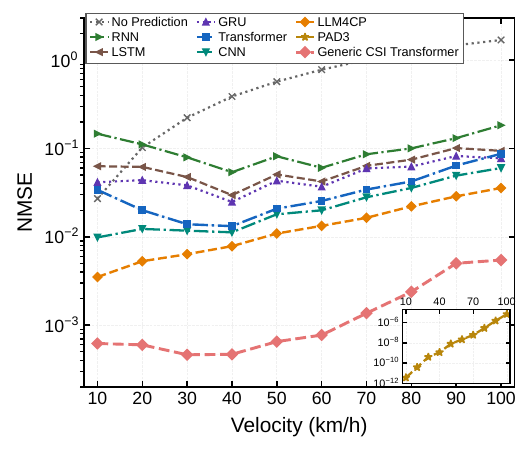}
    \caption{TDD NMSE across user velocities from $10$ to $100$~km/h under clean-input evaluation.}
    \label{fig:tdd_velocity}
\end{figure}

The corresponding TDD results in Fig.~\ref{fig:tdd_velocity} provide a same-band forecasting reference. PAD3 maintains the lowest prediction error across the velocity sweep, while the Generic CSI Transformer remains the strongest evaluated learning-based predictor.

\begin{figure}[!t]
    \centering
    \includegraphics[width=\columnwidth]{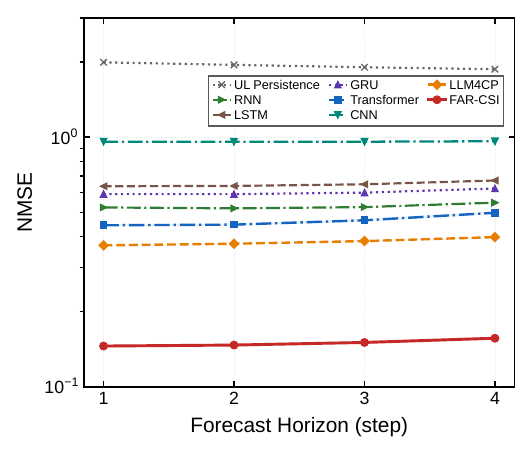}
    \caption{FDD NMSE versus forecast horizon for the four future prediction steps.}
    \label{fig:fdd_horizon}
\end{figure}

Figure~\ref{fig:fdd_horizon} examines whether the FDD prediction gain survives as the target moves farther beyond the observed interval. FAR-CSI retains the lowest error at all four future steps, showing that its advantage is not restricted to the nearest prediction instant but extends across the evaluated multi-step horizon.

\begin{figure}[!t]
    \centering
    \includegraphics[width=\columnwidth]{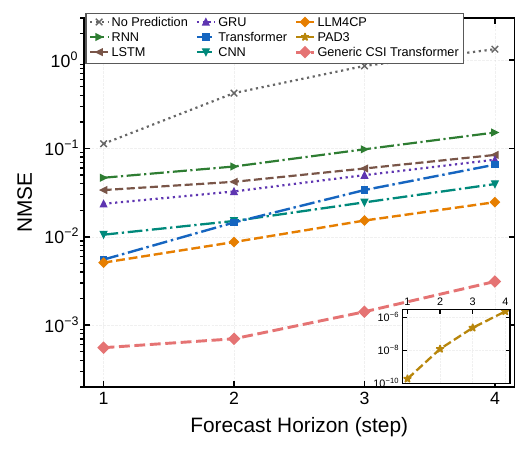}
    \caption{TDD NMSE versus forecast horizon for the four future prediction steps.}
    \label{fig:tdd_horizon}
\end{figure}

For the corresponding TDD horizon evaluation in Fig.~\ref{fig:tdd_horizon}, prediction error increases toward later future steps for the strongest same-band predictors. PAD3 remains the lowest-error reference, while the Generic CSI Transformer retains the strongest performance among the evaluated learning-based predictors.

\subsection{Data Efficiency}
\label{subsec:data_efficiency}

\begin{figure}[!t]
    \centering
    \includegraphics[width=\columnwidth]{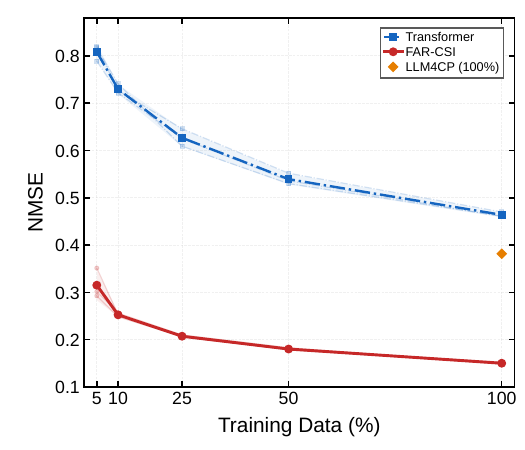}
    \caption{FDD NMSE under different fractions of downstream training data. FAR-CSI and Transformer are evaluated using the indicated training fractions, while LLM4CP is shown only as a separate full-data reference.}
    \label{fig:fdd_fewshot}
\end{figure}

Across all evaluated training fractions, FAR-CSI maintains lower NMSE than the reimplemented Transformer, as shown in Fig.~\ref{fig:fdd_fewshot}. Even with only $5\%$ of the downstream training data, FAR-CSI achieves lower NMSE than both the full-data Transformer and the separately evaluated full-data LLM4CP reference. The FDD forecasting advantage therefore persists under substantially reduced downstream supervision.

\subsection{Robustness to Noisy Historical CSI}
\label{subsec:noisy_csi}

\begin{figure}[!t]
    \centering
    \includegraphics[width=\columnwidth]{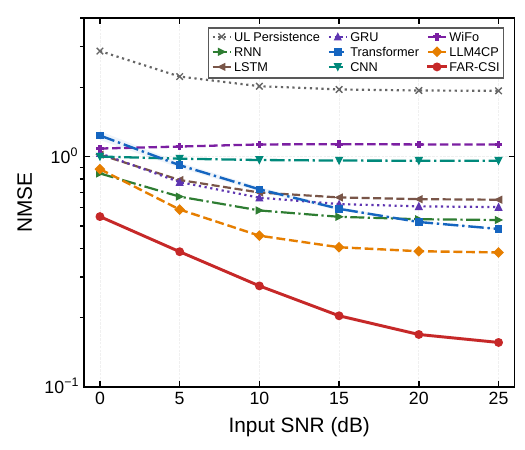}
    \caption{FDD NMSE versus historical-input SNR. Noise is applied only to the observed history, while prediction targets remain clean.}
    \label{fig:fdd_snr}
\end{figure}

With the trained predictors kept fixed, FAR-CSI maintains
the lowest NMSE among the evaluated FDD methods across
all tested finite-SNR conditions, as shown in
Fig.~\ref{fig:fdd_snr}. Its prediction error decreases
toward the clean-input level as the historical-input SNR
increases, with the accuracy advantage preserved despite
noise in the observed UL CSI.

\begin{figure}[!t]
    \centering
    \includegraphics[width=\columnwidth]{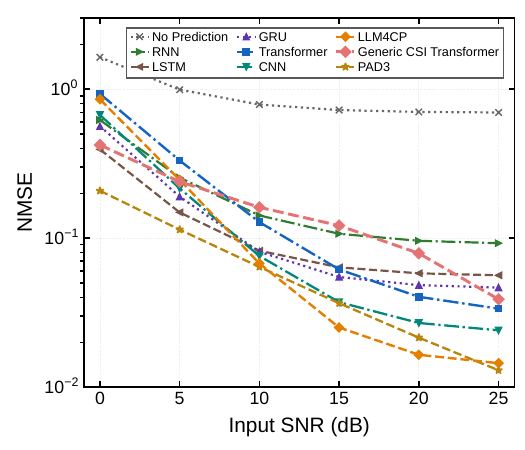}
    \caption{TDD NMSE versus historical-input SNR. Noise is applied only to the observed history, while prediction targets remain clean.}
    \label{fig:tdd_snr}
\end{figure}

The corresponding TDD results in Fig.~\ref{fig:tdd_snr} show that the lowest-NMSE method changes across the tested input-SNR levels.

\section{Conclusion}
\label{sec:conclusion}

FAR-CSI reframes FDD channel prediction as structured reconstruction over a joint temporal and operating-band representation. Its unified structured routing mechanism preserves this forecasting structure from contextual encoding through masked reconstruction to future-DL readout, allowing pretrained CSI modeling capacity to operate under the asymmetric historical-UL-to-future-DL setting. Under matched backbone and training conditions, FDD-aware routing achieves a mean paired NMSE reduction of $21.44\%$ relative to generic routing. Across the evaluated FDD baselines, FAR-CSI achieves the strongest overall prediction accuracy and MRT-based communication utility, while maintaining its advantage across different mobility levels, forecast horizons, reduced-data regimes, and noisy historical observations. These results support FDD-aware representation as an effective design principle for accurate and data-efficient cross-band channel prediction.
% ============================================================
% REFERENCES
% ============================================================

\bibliographystyle{IEEEtran}
\bibliography{references}

\end{document}